\documentclass[twocolumn,floatfix,preprintnumbers,aps,superscriptaddress]{revtex4}

\def\bea{\begin{eqnarray}}
\def\eea{\end{eqnarray}}
\def\ben{\begin{equation}}
\def\een{\end{equation}}
\def\benu{\begin{enumerate}}
\def\enu{\end{enumerate}}

\def\sss{\scriptscriptstyle\rm}

\def\1var{(\bx_1...\bx\N)}

\def\br{{\bf r}}

\def\bu{{\bf u}}
\def\bx{{x}}

\def\bj{{\bf j}}

\def\xc{_{\sss XC}}

\def\N{_{\sss N}}

\def\sph_int{ {\int d^3 r}}

\usepackage{graphicx}
\usepackage{psfrag}
\usepackage{amsmath}
\usepackage{bm}

\usepackage{epsfig}

\begin{document}
\title{An improved exchange-correlation potential for polarizability and dissociation in DFT}
\date{\today}
\author{Neepa T. Maitra}
\affiliation{Department of Physics and Astronomy, Hunter College and City University of New York, 695 Park Avenue, New York, NY 10021, USA}
\author{Meta van Faassen}
\affiliation{Department of Chemistry and Chemical Biology, Rutgers University, 610 Taylor Road, Piscataway, NJ 08854, USA}

\begin{abstract}
We propose a novel approach to the problem of polarizabilities and
dissociation in electric fields from the static limit of the
Vignale-Kohn (VK) functional. We consider the response to the purely
scalar part of the VK response potential. 
This potential has ground-state properties that notably
improve over the full VK response density and over usual (semi-)local
functionals. 
The correct qualitative behavior of our potentials means that it is
expected to work well for polarizabilities in cases such as the H$_2$ chain, 
and it will also correctly
dissociate open-shell fragments in a field.
\end{abstract}


\maketitle
Density functional theory (DFT)~\cite{HK64,KS65,K99b} has become the most
popular electronic structure method in a wide range of
problems in quantum chemistry, achieving an
unprecedented balance between accuracy and efficiency.  But in some
applications there are fundamental problems with standard
exchange-correlation (xc) approximations such as LDA and GGAs. The response to electric fields is severely overestimated for
long-range molecules and for molecular chains~\cite{CPGB98,GSGB99}.  Even for the
simple H$_2$ molecule dissociating in an electric field, LDA and GGAs yield unphysical fractional charges on each
atom~\cite{GB96,GSGB99}. This is a problem in many topical applications,
including molecular electronics~\cite{TFSB05,SZVV05} and nonlinear optics devices~\cite{KKP04,KRM94}.

It is now well known that the functionals in these cases must depend in an
ultranonlocal way on the density~\cite{GSGB99,MWY03,FBLB02}. 
There have been two
approaches: The first is exact exchange (EXX): this has
implicit non-local density-dependence through orbital-dependence~\cite{TS76}.
The second is Vignale-Kohn theory (VK)~\cite{VK96}, where a
linear response calculation is performed within time-dependent
current-DFT (TDCDFT), a natural extension of time-dependent-DFT~\cite{RG84}. 
VK uses functionals of the
current-density: the (time-integrated) current through a small volume
in space contains information about the density response far from that
volume, so functionals that are local in the current-density
are ultranonlocal in  the density. Both EXX and VK have
field-counteracting terms that decrease the response compared with
LDA/GGA, and so yield improved static polarizabilities for many polymer chains. An exception is the hydrogen chain, for which VK performs almost as poorly as LDA~\cite{FBLB02}.  Another problem that needs non-local functionals is
 the dissociation of simple diatomics in electric fields. As the
molecule is pulled apart, a field-counteracting step develops in the
exact Kohn-Sham (KS) potential midway between the atoms. In the limit of large
separation, the step size approaches a constant that realigns the highest
occupied molecular orbitals of the two atoms in the case of
two open-shell fragments, and vanishes in the case of two closed-shell
fragments. EXX captures the step in the latter, but until
now, no density functional approximation has captured the step in the former
case. VK has never before been applied to this problem.

  The calculations in~\cite{FBLB02} utilise the zero-frequency limit of
 VK, which contains additional ``dynamical'' xc fields, on top of LDA~\cite{UVb98}. The VK xc vector potential has both longitudinal as well as transverse components.  Static response
 calculations are technically outside the realm of validity of
 VK~\cite{VK96}, but its success in many such cases suggest such
 dynamical terms are also present in the true static
 functional  and contain essential physics~\cite{FBLB02,GV05}.

The fact that a transverse xc field persists in the static limit of VK means that, in contrast to usual response methods, caution must be used when interpreting 
the zero-frequency limit as  a ground-state perturbation.
Satisfaction of the adiabatic theorem requires that the response in the static
limit  is representable by a scalar KS
potential~\cite{GDP96}: For a perturbation turned on slowly enough, the system remains arbitrarily close to the 
 instantaneous ground-state. Functionals in any time-dependent theory should reduce to ground-state DFT ones, where xc effects are contained in a scalar potential. 
However, a 
  transverse field, with its non-zero curl, represents a non-conservative force, and so cannot correspond to a scalar potential underlying a conserved energy. 

In this paper we reconsider the VK response to electric fields from a new perspective.  
We first generate the self-consistent xc vector potential from a full VK
calculation in the static limit,  but then
discard its transverse component. Thereby we eliminate the non-conservative
part of the force.
We
gauge-transform the longitudinal part into a scalar potential which we view as a {\it ground-state xc response potential}.
This
approach is quite distinct from the previous use of VK~\cite{FBLB02},
where the key player is the {\it density response} of the VK
potential, $n_1^{VK}$.  This is not the ground-state
density response of the scalar potential above, 
because $n_1^{VK}$ is the full response to both the longitudinal and transverse fields of VK. Here, we 
consider the true ground-state response to just the scalar part of the
VK potential. 
We show that this ground-state VK potential has desirable features
arising from global field-counteracting terms.
The dissociation limit of the electron-pair bond
is correctly obtained, in contrast to the notorious fractional charges
that result from all previous density functional approximations, including LDA, GGA and
EXX~\cite{GB96,GSGB99,B01}.
For cases where the
 VK response density has not performed well, eg. hydrogen chains, features of this scalar VK potential suggest that our approach will perform very well.
Dynamical
terms that proved crucial in the usual VK approach to
polarizabilities, play an even more fundamental role here.

Consider a system initially in its field-free ground-state, of density
$n_0(\br)$.  In VK response theory, xc-contributions to a perturbative
time-dependent field are contained in a vector potential that is a
local functional of the induced current-density $\bj(\br
t)$~\cite{VK96}, ${\bf a}\xc[{\bf j}](\br t)$: in the frequency domain,
\ben
i\omega{\bf a}\xc[{\bf j}](\br,\omega) = \nabla v\xc^{\rm ALDA}(\br) - \nabla\cdot\tensor\sigma(\br, \omega)/n_0(\br)
\label{eq:axc}
\een
where, in the static limit, $\omega\to 0$, 
\ben
\sigma_{ij} = -\frac{3}{4i\omega}n_0(\br)^2f\xc^{\rm dyn}(\br)\left(\frac{\partial u_i}{\partial r_j} + \frac{\partial u_j}{\partial r_i} -\frac{2}{3}\nabla\cdot\bu \delta_{ij}\right)\,,
\label{eq:sigma}
\een
with $\bu(\br) = \bj(\br)/n_0(\br)$.
Here $f\xc^{\rm dyn}(\br)$ is a dynamical correction to the static scalar xc kernel of the homogeneous electron gas $f\xc^{\rm hom}$ in the long-wavelength limit, evaluated at the field-free density at $\br$, i.e. 
\ben
f\xc^{\rm dyn}(\br) = \lim_{\omega\to 0}f\xc^{\rm hom}(n_0(\br)) - \left.\frac{d^2 e\xc^{\rm hom}}{dn^2}\right\vert_{n_0(\br)}
\een

The {\it static} xc response of the homogeneous electron gas is given
by the second-derivative of the xc energy density $e\xc^{\rm hom}$. This 
differs from the zero frequency limit of the scalar kernel
by 
$f\xc^{\rm dyn} = 4\mu\xc^{\rm dyn}/3n^2$, where $\mu\xc^{\rm dyn}$ is the elastic shear modulus at $\omega\to 0$~\cite{QV02}. 
Thus VK contains dynamical corrections to ALDA that persist all the way through $\omega\to 0$. The response kernels are defined in VK by taking the wavevector $q$ of the
perturbing field to zero before taking the static limit, and these limits do not commute~\cite{GV05,UVb98,QV02,VK96}. 

We now define a ground-state potential from the static limit of the VK
response: Taking $\omega \to 0$, the longitudinal component of
Eq.~(\ref{eq:axc}) is gauge-transformed to a scalar potential,
$v\xc^{(1)}[n](\br)$, a non-local functional of the density:
$i\omega{\bf a}_{\sss XC,L}[{\bf j}](\br \omega) = \nabla
v\xc^{(1)}[n](\br)$.  We restrict now to cylindrically symmetric
linear systems, and {\it approximate} this potential in the following way. 
We consider Eqs.~(\ref{eq:axc}) and~(\ref{eq:sigma}) for a purely one-dimensional 
inhomogeneity~\cite{UVb98,SZVV05}, so we effectively average over the transverse directions in our cases.
We then obtain an approximate potential~\cite{UVb98,SZVV05} for linear systems:
\bea
\nonumber
v\xc^{(1)}(z)= f\xc^{\rm hom}(z)n_1(z)-\frac{n_0'(z)}{n_0(z)}f\xc^{\rm dyn}(z)\int_{-\infty}^{z} dz' n_1(z')\\
\nonumber
- \int_z^{\infty}dz'\frac{n_0'(z')}{n_0(z')}f\xc^{\rm dyn}(z')n_1(z') \\
+ \int_z^{\infty}dz'\left(\frac{n_0'(z')}{n_0(z')}\right)^2 f\xc^{\rm dyn}(z')\int_{-\infty}^{z'}dz''n_1(z'')
\label{eq:vk1d}
\eea
Here $n_0'(z) = dn_0(\br)/dz$ (where $\br = (0,0,z)$ is along the bond-axis) and $n_1(\br)=\nabla\cdot\bj/(i\omega)$ is the system's density response, which is taken in the zero-frequency limit.
This potential consists of four terms: the first is
the LDA response, local in the density response.
The second term is directly proportional to the local current-density
response. The third and fourth display global behavior
across the molecule, and are the key terms for the purposes of this paper.
It follows from the structure of the fourth term that any
polarization of the density, be it local or of charge-transfer nature,
yields field-counteracting behavior. The third term tends to align along the field, but is generally smaller than the fourth.

The first step is to run a zero-frequency VK response calculation on the chosen system in a weak external
field, $Ez$, placed along the bond-axis,
$z$. We utilize the ADF program package~\cite{ADF} with the TDCDFT
extension; see Refs.~\cite{FBLB02} for implementation details.
In the second step, the resulting density response is inserted in
Eq.~(\ref{eq:vk1d}) to define our ground-state potential. 
We study three classes of systems that are
challenging for usual semi-local density functionals: (i) dissociation
of the electron-pair bond (ii) a dimer composed of two closed-shell fragments at
large but finite separation, and (iii) a molecular chain. Atomic units
are used throughout.

{\it (i) Dissociation of the electron-pair bond: H$_2$-like
systems} The step that forms in the exact KS potential when a molecule composed of open-shell units dissociates
in an electric field prevents
dissociation to fractionally charged species, and is consistent with
the physical picture of two locally polarized species~\cite{GSGB99}. An
analogous step occurs in field-free dissociation of a heteroatomic
molecule composed of open-shell units~\cite{PPLB82,P85b,PL97c,AB85,GB96}. Its origin is static
correlation, and it is particularly difficult for
approximations to capture, eluding not only LDA/GGAs but also EXX. Ref.~\cite{B01} has a density-matrix solution for this. 

\begin{figure}[tbp]
\centerline{\psfig{file=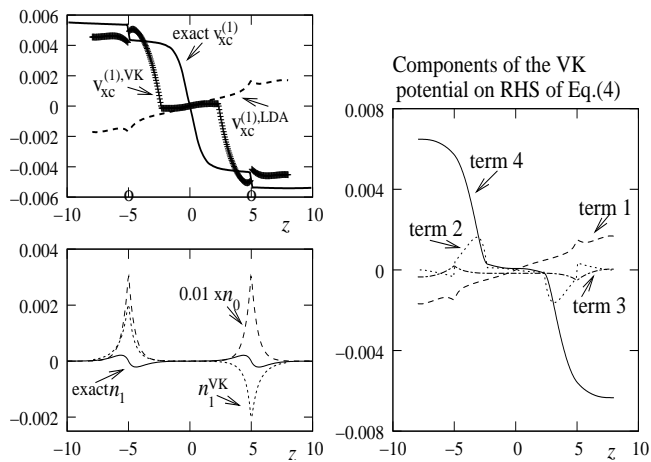,width=8.5cm,height=6cm}}
\caption{ H$_2$ at bond length 10 a.u., in a field of 0.001 a.u.
 The circles on the $z$-axis in the top panel represent the position of the atoms, as in all subsequent figures. A similar field-compensating behavior was observed in other diatomics composed of open-shell fragments, eg. Li$_2$. }
\label{fig:H2}
\end{figure}

The lower left panel of Figure~\ref{fig:H2} shows the field-free
density (scaled by 0.01), the exact density response within bound-state perturbation theory, and the VK
density response $n_1^{\rm VK}$ for H$_2$ at bond-length 10 au in a field of 0.001 au. The exact response demonstrates local
polarization with no charge-transfer, as expected. However, this is not
true of the VK (or LDA) response density, which yields fractionally charged
atoms. Correspondingly, the VK polarizability is grossly
overestimated, as it is in the LDA. The top panel of this same figure
shows the exact xc response potential $v\xc^{(1)}$ (within bound-state perturbation theory) and the
VK and LDA xc response potentials (i.e. subtracting the field-free
potential). The exact was obtained by numerically inverting the KS
equation, for the exact KS bonding-orbital composed of polarized
atomic orbitals. The salient feature of the exact $v\xc^{(1)}$ is the
field-counteracting step which compensates for the
difference in potential at the two separated atoms, and thus re-aligns
them when the total potential is considered. This step is 
missed by the LDA which has an along-field component, strengthening
the applied field. The step is also missing in EXX (not shown), as the
potential there is simply minus half the Hartree. The VK potential does capture the step. In the right panel
we see that the fourth term in Eq.~(\ref{eq:vk1d}) is responsible for
this. The VK potential falls a little short of the exact step, but
once the Hartree response potential
(field-counteracting~\cite{GSGB99}) is added to the VK xc 
potential, we expect that the total step completely compensates the change
in potential created by the external field.

The VK potential re-aligns the
two atoms, and so the ground-state of this system would
certainly not involve any charge transfer across the system, like in the exact case, in contrast to the VK density response (see Fig. 1). Indeed, as the bond-length $R$ increases, one finds that the
VK density globally polarizes even more, while the VK step size increases as
$ER$, maintaining the atoms at the same level. 
As the molecule dissociates, stretched H$_2$ has the metallic-like feature that its HOMO-LUMO gap vanishes. This may underlie the reason why VK exactly captures the step, since VK is based on the response of a metallic system, the weakly inhomogeneous electron gas.

It is important that the VK potential be evaluated
on the ``wrong'' density response to the full VK fields: if evaluated on
the exact response density, the field-counteracting term is reduced by
a factor of about 20.

\begin{figure}
  \centerline{\psfig{file=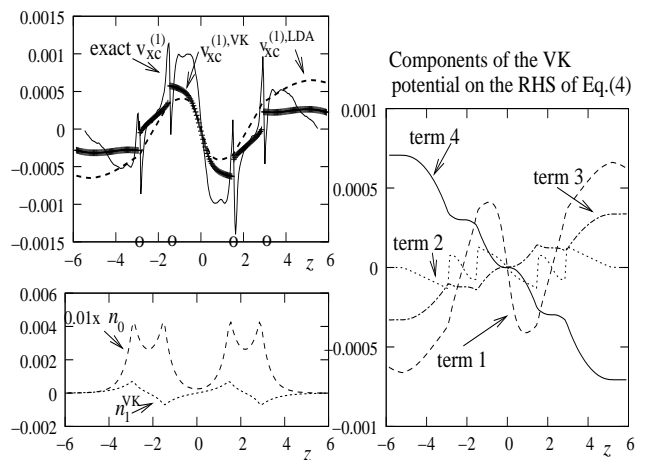,width=8.5cm,height=6cm}}
\caption{The H$_2$-dimer, with  H$_2$-- H$_2$ distance 3 a.u., and ``intramolecular'' H--H bondlength 1.4 a.u, in a 0.001 au field.}
\label{fig:2H2}
\end{figure}

{\it (ii) Dimer of two closed-shell units} Field-counteracting
behavior also arises from the exchange interaction between closed
shells, as explained in Refs.~\cite{GSGB99,GB01}. In contrast to the
step of (i), EXX methods can retrieve this
step~\cite{GSGB99}. In Figure~\ref{fig:2H2} we see that the
step is also nicely reproduced by the VK potential, shown in the top
panel. Also shown there is the exact xc response potential, calculated
from highly accurate wavefunction techniques in
Ref.~\cite{GSGB99}. 
 Although details of the
exact potential are missing in the VK response potential, the step is
clearly captured, with the correct magnitude of
drop in the potential between the up- and down-field molecules. In
contrast, the LDA or any GGA has no net drop.

The right panel shows that the net step in the VK potential
is a result of competition between the third and fourth terms. Note
that in this case there is largely local polarization in the
molecules, with small charge transfer (hardly discernable in the lower left panel).

The VK density response and polarizability are again very close to the LDA
ones, both overestimating the exact polarizability. However, the ground-state response using the VK
potential shown, is expected to yield polarizabilities much closer to the
exact one, because of the field-counteracting behavior.

The magnitude
of the exact step between two closed-shell units
decreases as the intermolecular separation increases~\cite{GSGB99}. This is
{\it not} the case for the VK step asymptotically.
Essentially, the VK step arises from integrals over the
current-density (fourth term in Eq.~(\ref{eq:vk1d})); this integral persists at large separation, due to the local polarization.

\begin{figure}[tbp]
  \centerline{\psfig{file=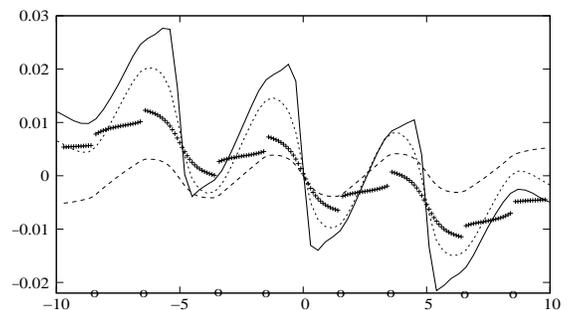,width=7.3cm,height=4cm}}
\caption{H$_2$ chain in a field of 0.005au: OEP (solid), KLI (dotted), LDA (dashed), and VK potentials (+). The H--H distances alternate between 2 and 3 a.u. }
\label{fig:H2chain}
\end{figure}
{\it (iii) Long-chain molecules} Polarizabilities and
hyperpolarizabilities of long-chain polymers are notoriously
overestimated in LDA/GGA calculations. Recent work has shed much light
on the role of ultranonlocal density dependence in these systems. It
has been demonstrated that orbital functionals such as EXX
within OEP can significantly improve the
polarizabilities~\cite{KKP04,GSGB99, MWY03}, due to field-counteracting
behavior in the exchange potential.  Approximate exchange methods such
as KLI~\cite{KLI92} also give corrections over LDA, although not
as strongly as full exchange~\cite{KKP04}. 
For example, for the
hydrogen chain in Figure~\ref{fig:H2chain}, the LDA, KLI, and OEP
polarizabilities are 114.6, 90.6, and 84.2 a.u, respectively. 
The VK functional was also applied~\cite{FBLB02} and gave 110.20 a.u, hardly an improvement
over LDA; in contrast to its dramatic improvement for many other molecular chains.
Again, the scalar VK {\it potential} has field-counteracting behavior, as shown in Figure~\ref{fig:H2chain}. This potential will give improved results (comparable to KLI)
 when used in a computation of
ground-state response and polarizabilities.

In Ref.~\cite{KKP04}, the importance of intermolecular barriers in the
{\it field-free} ground-state potential was stressed: these are
lacking in  LDA, underestimated in KLI, but captured well in
OEP. Although formal arguments point to LDA as the correct
field-free potential to be used in conjunction with the VK response
potential, it will be interesting to compare 
polarizabilities using the VK response potential on top of a
field-free OEP potential.

In summary, we have shown that the scalar part of the VK response
 potential contains crucial field-counteracting terms making it a
 good candidate for response properties of
 effectively one-dimensional systems in electric fields. It is the
 first approximation that captures the step in a molecule composed of open-shell fragments dissociating in an
 electric field. It captures field-counteracting terms in
 systems of two or more closed-shell units, and is promising for
 long-chains, being numerically less intensive than EXX methods which
 also have been successful for these problems. The scalar VK potential works
 well even when the full VK response does not!
In all cases studied, $n_1^{VK}$ is very close to the LDA response, suggesting that the effects of the transverse VK and dynamical-longitudinal fields
somewhat cancel.
Satisfaction of the adiabatic theorem means that the transverse component
of the response field should vanish in the static limit, suggesting that VK should be corrected by dropping its transverse part.
This supports our results, where the ground-state response to the longitudinal component was considered.
An alternative choice is to discard the transverse field from the start, and consider the
self-consistent response to the purely longitudinal field.  
The results are not as good:
the field-counteracting nature of
Eq.~(\ref{eq:vk1d}) is most effective when evaluated on the full (wrong)
density-response $n_1^{VK}$ rather than that obtained
self-consistently. 
The non-self-consistent aspect of our present approach is somewhat
dissatisfying from a rigorous viewpoint; on the other hand, there are many other
situations in DFT where such post-self-consistent approaches have  success eg. Ref.~\cite{PKZB99}.

Finally, we note that the VK response
 potential smears over some local details of the true response potential
 (see e.g. Fig.~\ref{fig:2H2}). Work is underway to
 investigate if this has a significant effect on the global polarizability.

We thank S. K\"ummel for his OEP and KLI data on the H$_2$ chain, and
K. Burke and G. Vignale for fruitful discussions. This work is
financially supported by the American Chemical Society's Petroleum
Research Fund and National Science Foundation Career CHE-0547913.

\end{document}